\documentclass{elsart}

\usepackage{graphicx,epsfig,graphics}
\usepackage{amsmath,amssymb}
\usepackage{bm}
\usepackage{rotating,texdraw}
\usepackage{color}
\usepackage{dsfont}

\newcommand {\be} {\begin{equation}}
\newcommand {\ee} {\end{equation}}
\newcommand {\fig} {Fig. \ref}

\newcommand {\bq} {\begin{eqnarray}}
\newcommand {\eq} {\end{eqnarray}}

\DeclareMathOperator \ash {ash}
\DeclareMathOperator \ach {ach}

\begin{document}
\begin{frontmatter}
\title{Depinning of a discrete elastic string
from a two dimensional random array of weak pinning points}

\author{Laurent Proville}
\address{CEA, DEN Service de Recherches de
M\'etallurgie Physique, \\
        F-91191 Gif-sur-Yvette, France}

\date{\today}

\begin{abstract}
The present work is essentially concerned with
the development of statistical theory for the low temperature
dislocation glide in concentrated solid
solutions where atom-sized obstacles impede plastic flow.
In connection with such a problem, we compute
analytically the external force required to drag an elastic string
along a discrete two-dimensional square lattice,
where some obstacles have been randomly distributed.
The corresponding numerical simulations
allow us to demonstrate a remarkable
agreement between simulations and theory for
an obstacle density ranging from 1 to 50 $\%$
and for lattices with different aspect ratios.
The theory proves efficient on
the condition that the obstacle-chain interaction
remains sufficiently weak compared to the string stiffness.
\end{abstract}

\end{frontmatter}

\maketitle

\section{From the solid solution strengthening theory}
The statistical theory for solid solution hardening (SSH)
emerged from the seminal works of Sir N. Mott\cite{Mott1952} and
his near colleagues, F.R.N. Nabarro\cite{Mott1948,Nabarro1985} and
J. Friedel \cite{Friedel1964}.
The early analytical theory,
perfected and extended by other contributors, as for instance R.
Fleischer, R. Labusch and T. Suzuki
\cite{Labusch1972,Fleischer1963,Fleischer1964,FleischerHibbard1963,Suzuki1991}
applies to substitutional alloys where the solute atoms can be
considered as immobile during the dislocation glide, by contrast to
the cases where dislocations may drag along an atmosphere
of fast diffusing impurities.
In face centered cubic (fcc) alloys,
the critical resolved
shear stress (CRSS) was then expected to increase in proportion to $c_s^{\eta}$ with
$c_s$ as the atomic concentration of solute atoms and $\eta$ as an exponent
depending on the assumptions made on the interaction between
dislocations and foreign atoms: $\eta=1/2$
in Friedel-Fleischer (FF) theory while $\eta={2/3}$ in Mott-Nabarro-Labusch (MNL) theory
and $\eta=1$ in Friedel-Mott-Suzuki (FMS) \cite{Friedel1964,Suzuki1991}.
Within analytical theory for SSH, the dislocation is thought of
as a continuous elastic string impinged
on a two-dimensional (2D) random static potential.
The depinning transition in such a model is a
typical issue of statistical physics, belonging to a broad
class of problems concerned with extended
interfaces motion in heterogeneous materials\cite{Zapperri,DeGennes1984,Vinokur1994,Brazovskii2004,Vinokur2006,Bouchaud1997,Fisher,Fisher1993,LeDoussal2000,Wiese,Tanguy2004,Vinokur2002,Rosso2007}.

The recent developments of three-dimensional atomistic
simulations (3D-AS) allowed to work on more realistic models for dislocations in solid solutions\cite{Rodary2004,Olmsted2005,Proville2006,Tapasa2006,Tapasa2007,Patinet2008,Marian2006,Curtin2006}.
Though 3D-AS confirmed that a large part of the dislocation pinning hinges on
the impurities situated in the crystal planes that bounds the
dislocation glide plane\cite{Rodary2004}, the simulations revealed also the complexity of
the dislocation-obstacle interaction. In fcc alloys,
the geometry of the dislocation core, dissociated in two Shockley partials separated by a $(111)$ stacking fault ribbon
undermines the simple picture of an elastic line in interaction with a single type of obstacles, as stated
in the basic version of SSH theory. Instead,
the pinning forces differ according to partials and to the obstacle positions, i.e., above or below the glide plane\cite{Proville2006,Patinet2008}.

On the other hand, the nanometric scale of the atomistic simulations,
a stringent limit imposed
by the computational load, hinders the direct extrapolation
of simulation results to macroscopic samples.
A multi-scale approach is thence required to link the atomistic studies to the
realm of materials sciences. A manner to proceed consists in
incorporating some of the atomic
ingredients revealed in 3D-AS
to a discrete version of the elastic string model
that remains tractable even for large dimensions \cite{Madec2003}.
The discretization of the string model should allow to transfer
of the atomic details.

In the present paper,
the impact of the discretization on the depinning transition
is analyzed thoroughly. The elastic string
is replaced with a discrete spring chain, the nodes of which move
on a 2D square lattice and interact with
some pinning points randomly distributed on lattice sites.
This very simple model allows us to devise an analytical theory
which accounts for the discreteness of
the obstacle distribution and thus opens a promising perspective
to integrate more of the atomic details.
In order to demonstrate the accuracy of the theory,
we compute directly the critical
external force within numerical simulations applied to the
discrete string model. Theory and simulations
agree remarkably well for a broad range
of model parameters, e.g., (i) the in-plane obstacle
density $c_s$, (ii) the lattice size in every direction of space, (iii) the maximum pinning force $f_M$
and (iv) the potential interaction
cutoff $w$
that characterizes the obstacles. The
theoretical predictions are proved reliable on the condition that
$f_M$ and $w$ remain
smaller than certain bounds varying with $c_s$.
For a dense distribution of weak pinning points, the critical configuration
of the chain is found to be a quasi-straight line parallel to the atomic rows.
The depinning is then shown to occur at some vacant site clusters (VSC) which the typical
size is explicitly related to the lattice dimensions in both directions of space. The
external force required to drag along the
spring chain over a finite distance reflects such a size dependence.
Noteworthily the effective density exponent $\eta$
is also found to vary with lattice
dimensions, in contrast with expectations drawn on standard SSH theory.

Our report is organized as follows. In Sec. II, the spring chain model
is introduced and the direct numerical computations are described. In Sec. III,
the statistical theory is derived and compared to the numerical data.
The results are resumed and commented in Sec. IV.

\section{The phenomenological spring chain model \label{Sec:simul}
}
The model
proposed hereafter belongs to the wide class of elastic interface models,
extensively studied in statistical physics \cite{
Fisher,Fisher1993,LeDoussal2000,Wiese,Tanguy2004,Vinokur2002,Rosso2007}. A one dimensional elastic string
is discretized with a spatial step $b$,
equivalent to the shortest interatomic distance in solids. Each node
of the discrete chain is bound to its first neighbor by an
harmonic spring of strength $\Gamma$.
The two
quantities, $b$ and $\Gamma$ are chosen to scale distances and forces, respectively.
The
spring chain nodes move along the column
of a square lattice. The size of the lattice
in the direction of the chain is denoted as
$L_y$ whereas the distance over which
the chain is dragged is $L_x$.
The 2D random array
of obstacles is constructed by selecting the occupied lattice sites, up
to a number of obstacle equals to $c_s L_x L_y$, where $c_s$
is the obstacle density.
Since the depinning process
occurs when the chain nodes pass the force maximum,
the interaction potential is expended as a polynomial function in the vicinity
of such a maximum.
Assuming that the interaction
is attractive and that the potential is
symmetric with respect to its minimum, we obtain
a polynomial function of at least fourth order:
\begin{eqnarray}
V(x)&=&V_0 ({x^2}/{w^2}-1)^2\ \text{for}\ |x|<w\nonumber\\
V(x)&=&0\ \text{for}\ |x|>w,
\end{eqnarray}
which corresponds to a force
$f(x)=-4V_0 ({x^2}/{w^2}-1){x}/{w^2}$,
with a maximum value $f_M= 8 |V_0| /(3\sqrt{3} w)$,
attained when $x=\pm w/\sqrt{3}$. The chain nodes interact
solely with
obstacles situated in the column along which they may glide. The polynomial force
with a distance cutoff $w$ is obviously very far from the dislocation-solute interaction, characterized
by a decrease of Coulomb type.
Hereby we describe only the local potential variation yielded when a solute atom visits
a dislocation core. The parameter $w$ fixes how the interaction decreases in the vicinity of the force maximum. Both $f_M$ and $w$ can be extracted from atomistic data as those reported in
\cite{Proville2006,Patinet2008}.

The dimensionless
over-damped Langevin dynamics for the chain node $k$ is given by:
\be\label{Langevin}
\dot{x_k}= [x_{k+1}+x_{k-1}-2x_{k}] + \tau
- \sum_i 4 V_0 \frac{(x_k-s_{k,i})}{w^2}
(\frac{(x_k-s_{k,i})^2}{w^2}-1),
\ee
where $x_k$ is the position of the node $k$, $\tau$ is the
external force and
$s_{k,i}$ is the coordinate of the i{\it th} obstacle in the k{\it th}
row. For the weak pinning forces we are concerned with, the chain
strain remains very small such that the anharmonic terms in the spring tension
have been neglected. Properly scaled,
the continuous version of the spring chain model served in the
development of the
SSH theory \cite{Mott1952,Nabarro1985,Friedel1964,FleischerHibbard1963}.

In the direct numerical simulations of Eq. \ref{Langevin}, $\tau$ is incremented
adiabatically in the course of the integration of the
chain motion. Once [$\sup_k|\dot{x_k}|$] is
inferior to a certain precision (i.e., $10^{-7}$) the external force is incremented. Before
each increment, the chain configuration is recorded and once the
chain has run over a distance $L_x$, the integration is stopped. The latest anchored
configuration corresponds to the strongest one and the
associated external force is denoted as $\tau_c$, i.e., the static depinning
threshold. We performed this
type of simulations for different lattice aspect ratios, varying $L_x$ and $L_y$ and for
different obstacle densities ranging from 1 to 50 $\%$.
Various algorithms for the random numbers generator needed
to build the obstacle array were tested and
no significant difference was noticed in the end results.

\begin{figure}
\includegraphics[width=8.cm]{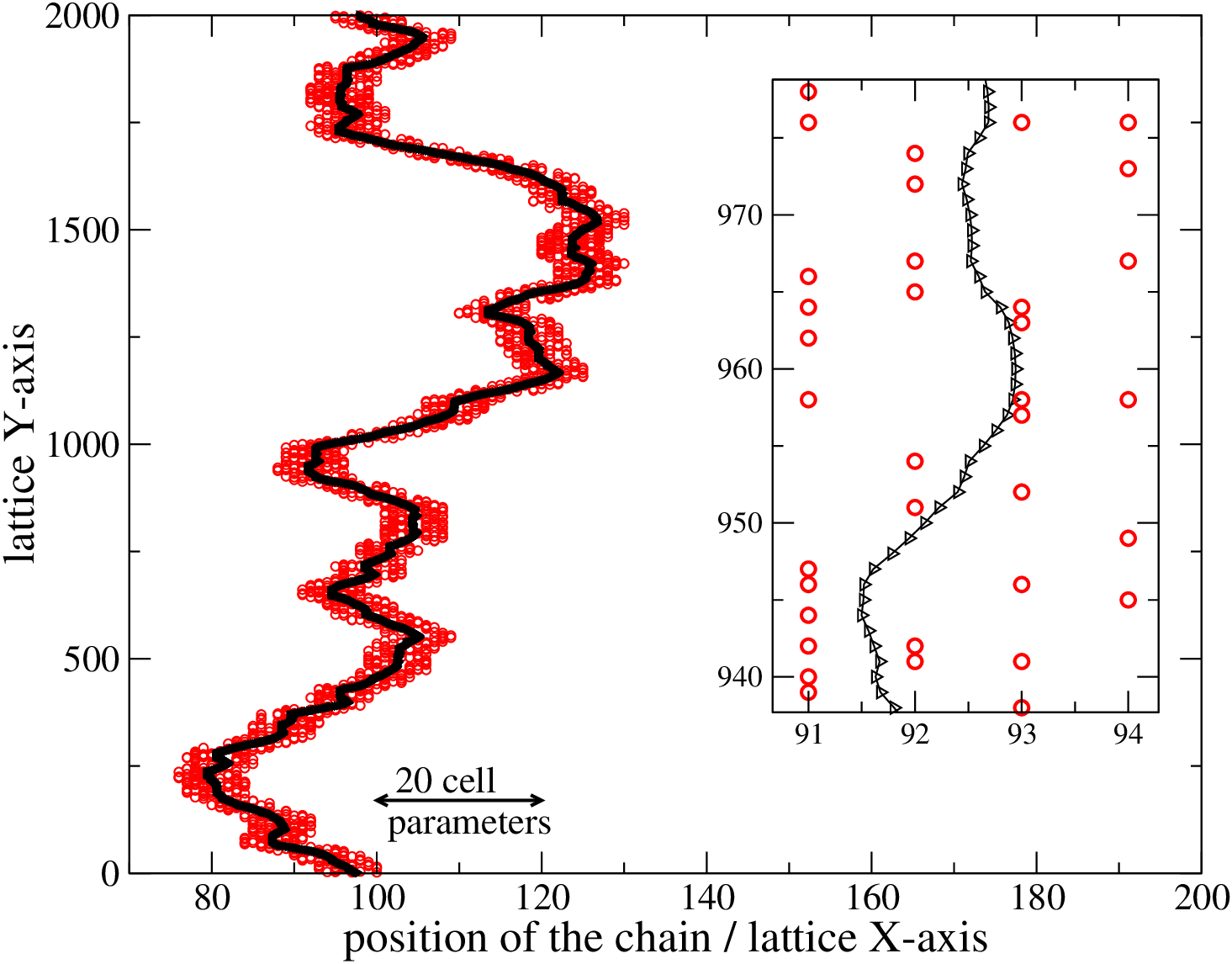}
\caption{\label{fig:FM0_1} \scriptsize  (color online)
Strongest pinning configuration of the spring chain on a
random array of obstacles (circles) for $f_M=0.1$, $L_y=2000$, $L_x=500$, a density
$c_s=16$ $\%$ and an interaction cutoff $w=1$. Only obstacles close from the chain
have been reported for clarity. Inset shows a magnification of obstacles (circles) and nodes (triangles) of a chain segment. X and Y axis have
different scaling for convenience of the plot.}
\end{figure}
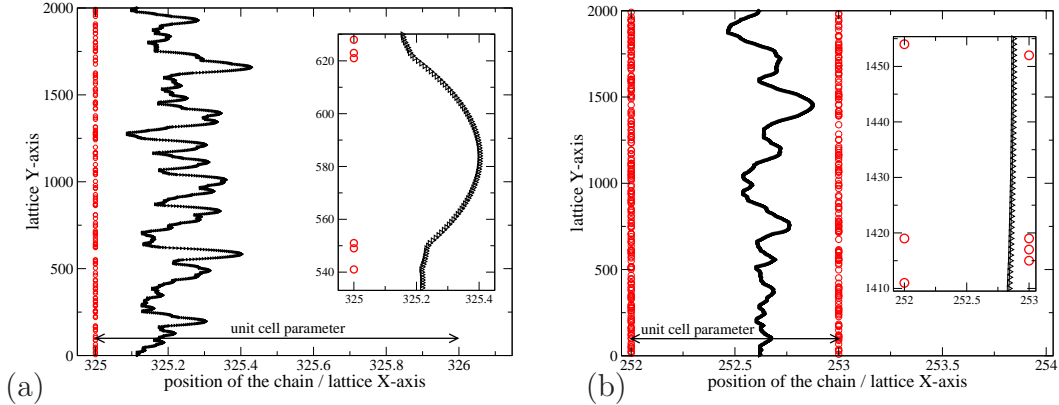
\begin{figure}
\noindent
\includegraphics[width=6.5cm]{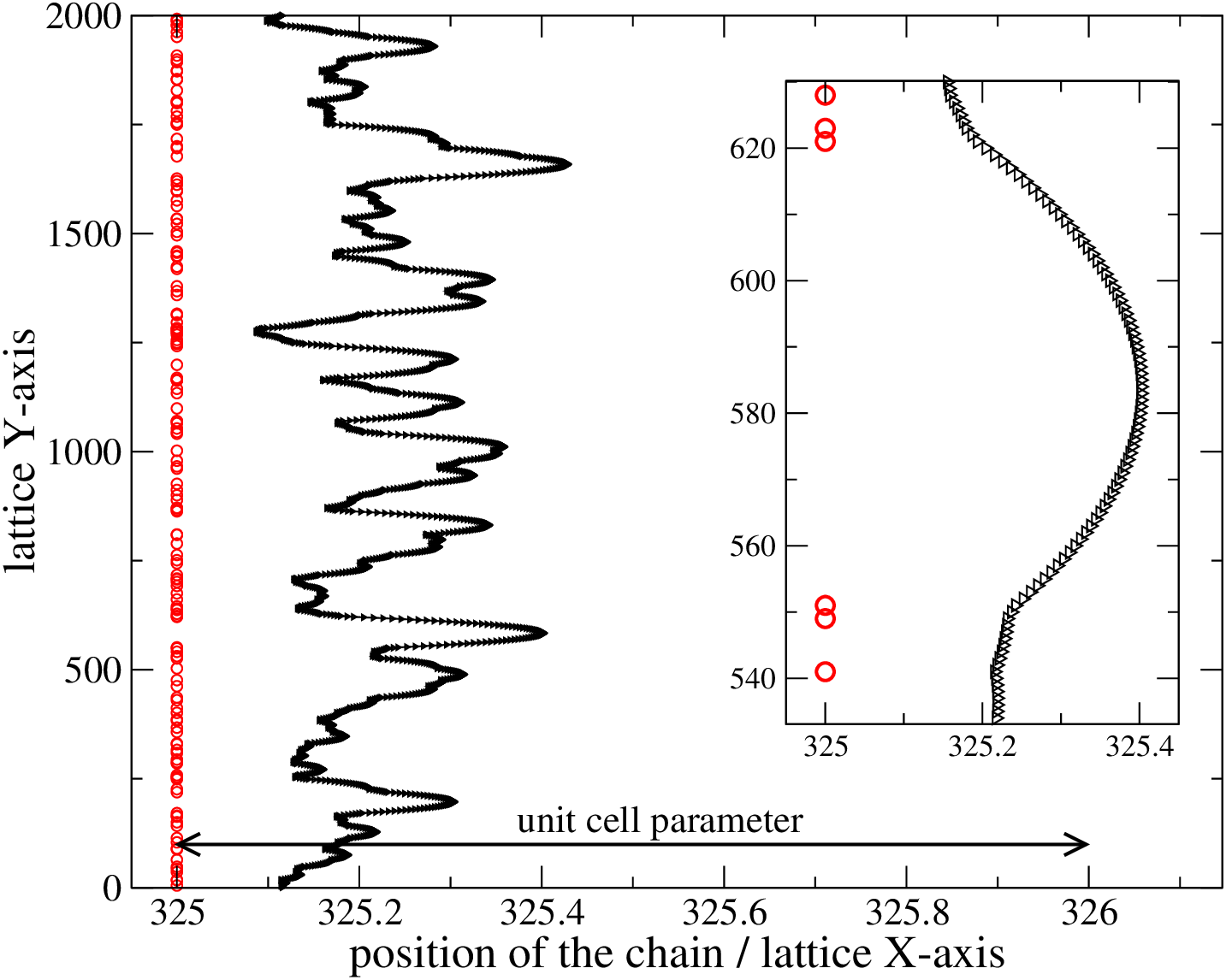}\hspace{.6cm}
\includegraphics[width=6.5cm]{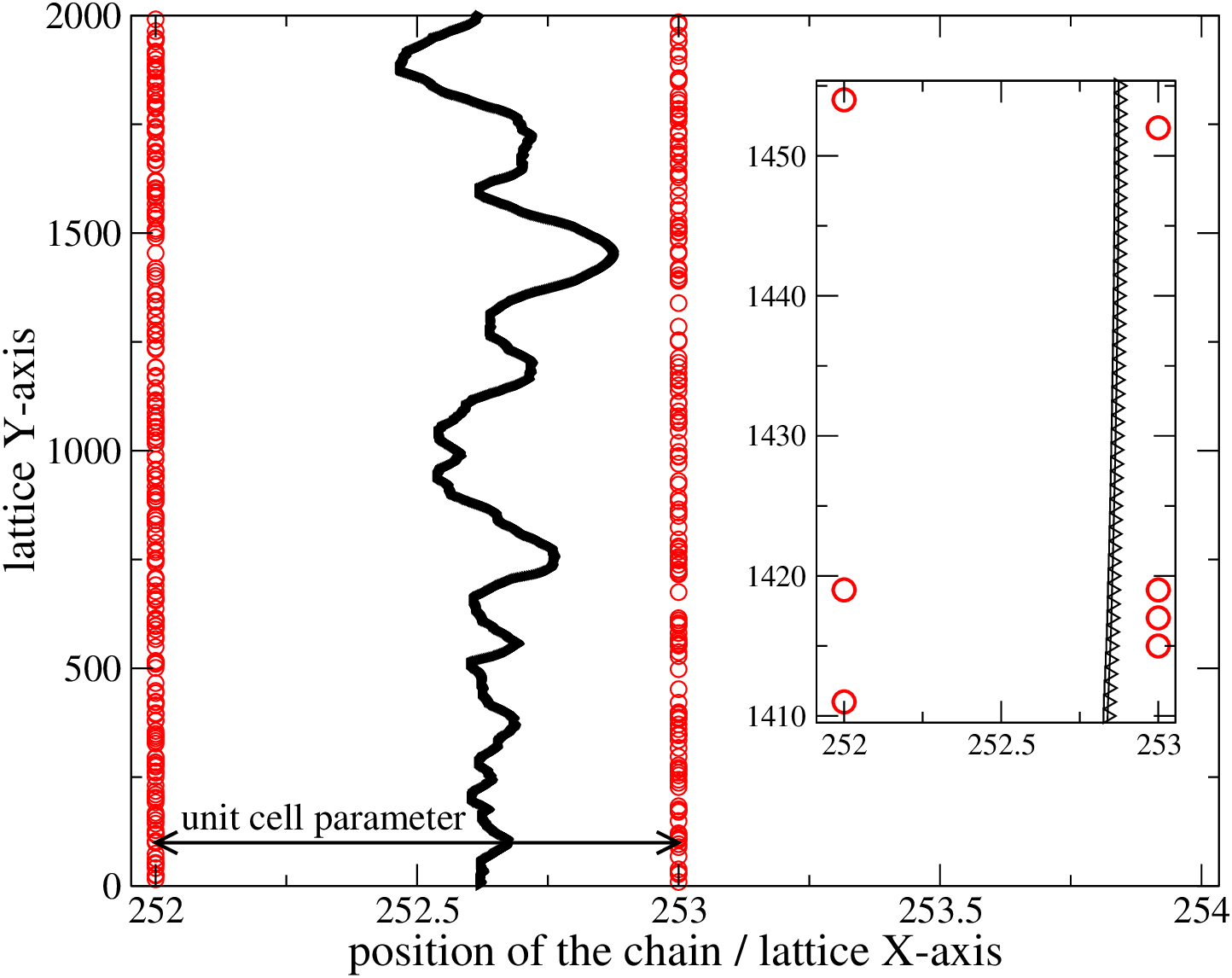}
\caption{\label{fLP3} \scriptsize (color online) Strongest
pinning configuration on a
random array of obstacles (circles) for a chain length $L_y=2000$ and
a drag distance $L_x=500$. In (a)
the interaction cutoff is $w=0.5$, the obstacle
strength $f_M=0.005$ and the obstacle density
$c_s=7$ $\%$.  In (b) $w=1$, $f_M=0.001$ and $c_s=9$ $\%$. In the
insets some segments are magnified with chain nodes marked
as triangles (not visible on the main graphics). The X and Y axis
scales differ for convenience of the
plot.}
\begin{picture}(100,10)(0,0)
\put(0,92){\makebox(0,0){(a)}}
\put(220,92){\makebox(0,0){(b)}}
\end{picture}
\end{figure}

In Fig. \ref{fig:FM0_1}, we report the strongest chain configuration,
obtained from the numerical simulations
for a pinning strength $f_M=0.1$. The critical profile
is found to be wavy and to cross at least 40 lattice rows.
In Figs. \ref{fLP3} (a) and (b), the critical chain profile is shown
for smaller values of $f_M$, i.e., two orders of magnitude smaller than
the one used in Fig. \ref{fig:FM0_1}. We note that the entire string length is
bounded by only two rows. The simulations evidence actually a well known
feature for pinning of extended defects, namely weaker the obstacles
flatter the shape of the critical configuration.
A perfectly rigid string would even experience a null force since then $f_M$ would be
negligible compared to the spring tension.
However, as soon as some elasticity enters into play, the pinning strength becomes positive.
The present work is essentially concerned with cases like those presented in
Figs. \ref{fLP3} (a) and (b), where the elastic string shape is quasi-straight.
In such situations,
the string roughness is inferior or of the order of the inter-atomic spacing.
The result shown in Fig. \ref{fig:FM0_1} only served us for comparison in order to introduce
our problem. Such a case of wavy critical profile
has been studied extensively,
both through numerical simulations\cite{Foreman,Rodney2006,Picu2007,bestPhysicist,Arsenault1989}
and analytical works \cite{Labusch1972,Fleischer1963,Fleischer1964,FleischerHibbard1963}.

In the insets shown in  Figs. \ref{fLP3} (a) and (b), it is worth noticing that
along the rows that bound the spring chain, some holes appear in the obstacle distributions.
Hereafter,
we dubbed such holes {\it vacant
site clusters}. The sampling of such density fluctuations along lattice rows
plays a key role in the determination of the critical drag force.

\section{Vacant site cluster sampling theory}
\label{HS}
\subsection{The tightly bound chain}
\label{Sec:HS_A}
In \fig{fLP3} (a), we noticed that for some parameters the
strongest configuration of the
spring chain remains tightly bound to the single lattice row at the back of the chain, i.e., most of the chain nodes
are closer from the back row than from the next nearest one and thereby
the chain does not cross several lattice rows.
In the present section, a theory is devised to compute the
critical drag force corresponding to critical configurations
like the one shown in Fig. \ref{fLP3} (a).
Further, the theory will be extended to
cases with broader interaction cutoff.

When the chain is tightly bound to the back
row, the string can be viewed as quasi-straight, notwithstanding
the bulges formed between rows. When $w\leq 0.5$, we can assume
that the chain interacts with rows one by one and it is natural to
work on the hypothesis that for such a system the
strength of the random lattice is fixed by its more crowded row.
To translate such a
remark into some algebra, one needs to study the
sampling of obstacles on a finite size lattice $L_x\times L_y$. We notice that
the purely random planar distribution follows
Bernoulli's binomial law and the
number of obstacles $N_o$ involved into a single row of length $L_y$ is
then a random variable which probability is given by:
\be\label{Eq:Binomial}
\rho(N_o)= \mathds{C}_{N_o}^{L_y} \ c_s^{N_o}\ (1-c_s)^{L_y-N_o},
\ee
where $\mathds{C}_{N_o}^{L_y}={L_y !}/{N_o! (L_y - N_o)!}$.
Such a statistical distribution can be approximated with
a Poisson law in the limit of large $L_y$. However
such a rounding yields some error for smallest $L_y$
we are concerned with, so we keep the binomial
formulation of Eq. \ref{Eq:Binomial}.
The probability for a row to involve less than $N$ obstacles is
$\sum_{N_o<N} \rho(N_o)$ and therefore in a
set of $L_x$ rows
the probability for having a row with $N_m$ obstacles and
$(L_x-1)$ rows with a number of obstacles inferior to $N_m$ is :
\be\label{eq:beta}
\beta(N_m) = L_x \ [\rho(N_m)]\ [\sum_{N_o<N_m} \rho(N_o)]^{L_x-1}.
\ee
The maximum number $N_m$ fixing the number of obstacles in the
denser row depends only on the lattice dimensions in each direction of space
and on the overall obstacle
density $c_s$. It is easily computed numerically, paying attention
to avoid overflows in factorials computation.
The mean density in the denser row is
then $c_m=N_m/L_y$.
When an excess of
vacant sites emerges at some place along the denser row, such segment is
weaker than others where the obstacles are more crowded. Thence the
spring chain should start the crossing
at the largest vacant site clusters (VSC). The typical size of
such VSC must now be determined.
Actually the mean number of VSC in a row which obstacle density is fixed to $c_m$
is $L_{VSC}=(c_m L_y -1)\approx N_m$. The normalized probability to find
a VSC with exactly n vacant sites is $c_m (1-c_m)^{n}$ while
the probability for a VSC which size is inferior to $n$
is $[1-(1-c_m)^n]$. The probability
to find a VSC of size n and ($L_{VSC}-1$) VSC with size inferior to n is proportional to:
\be\label{eq:gamma}
\gamma(n)= L_{VSC} \ [c_m (1-c_m)^n ] \ [1-(1-c_m)^n ]^{L_{VSC}-1}.
\ee
The mean size of the largest VSC in the denser row is thus:
\be
n_v = \sum_n [n\gamma(n)]/\sum_n \gamma(n).\label{trans}
\ee
Such a maximum VSC is surrounded by other VSC's that mean size
is given by: $[\sum_{n<n_v} n c_m (1-c_m)^n ]/[\sum_{n<n_v} c_m (1-c_m)^n]$
which for convenience is denoted as $(m-1)$ with:
\be
m=\frac{1}{c_m} - n_v \frac{(1-c_m)^{n_v}}{1-(1-c_m)^{n_v}} \label{eq:meanSPACING}.
\ee

\begin{figure}
\includegraphics[width= 8cm]{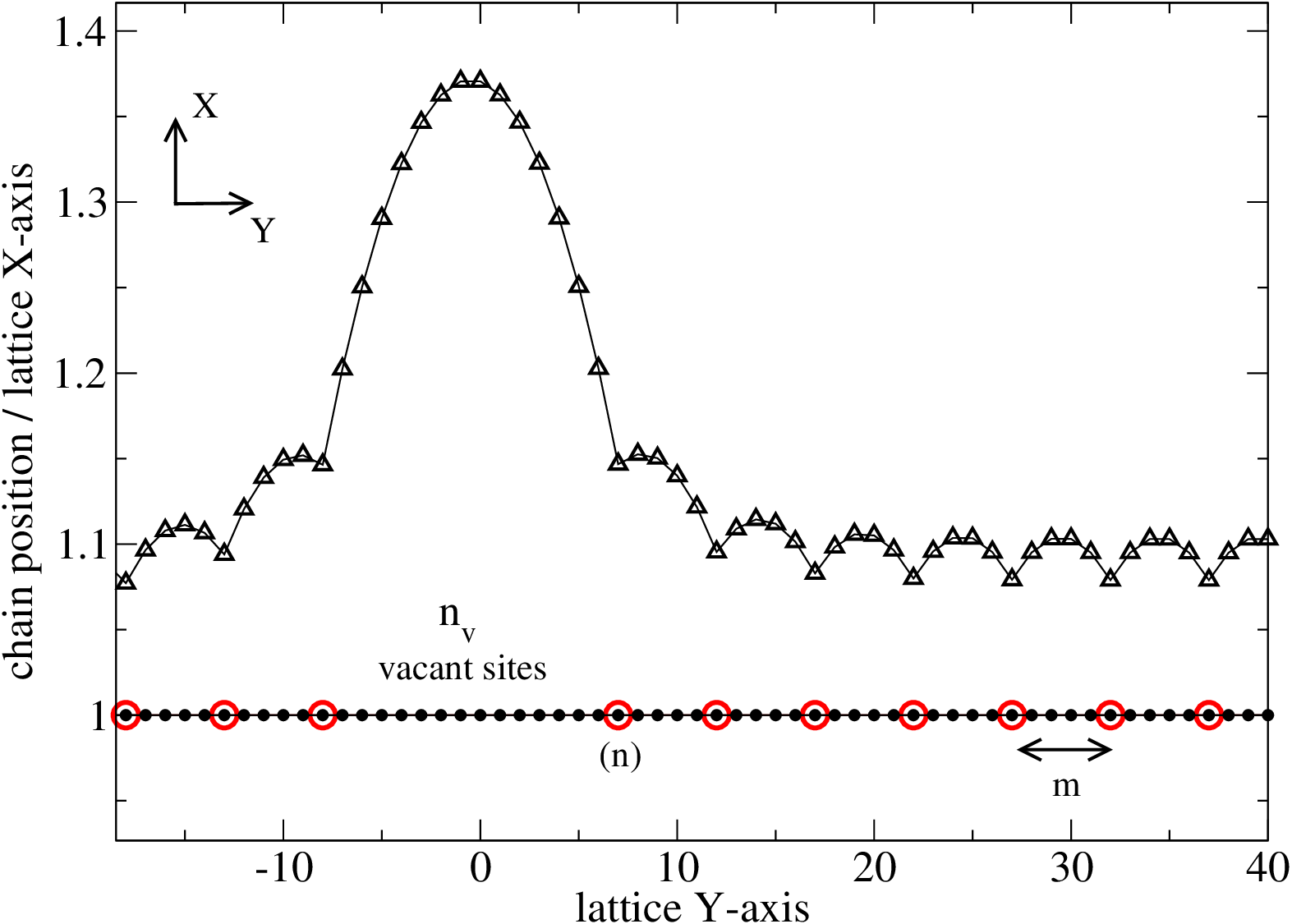}
\caption{\label{fLP1} \scriptsize (color online) Schematic representation of the
model used for a quasi-straight spring chain tightly
bound to a single lattice row. The small full circles represent
the lattice sites, the large open circles represent the obstacles
and the triangles are for the spring chain nodes. The average number $n_v$ of
vacant sites involved in the largest vacant site cluster is determined through
Eq. \ref{trans} and Eq. \ref{eq:gamma}.
The average spacing $m$
between the obstacles on both sides of the largest
vacant site cluster is fixed by Eq. \ref{eq:meanSPACING}.}
\end{figure}

To compute the external force associated with the strongest
binding row we consider the segment of $n_v$ vacant
sites as embedded into a regular lattice of obstacles spaced by a
mean distance $m$. Such a
mean-field construction is illustrated within Fig. \ref{fLP1},
where spring chain's nodes (triangles) are bound to the lattice
sites occupied by the obstacles (large open circles).
The array of
obstacles is assumed to be centro-symmetric, so we ascribe the
label 0 to the center of symmetry which corresponds to the top of the bulge.
We also define a new variable $n=(1+n_v)/2$ for convenience of the notations. Under the
external applied force $\tau$, The
force balance sheet, for say the left hand side
of the chain leads to the set of equations:
\begin{eqnarray}
  F_0  &=& -\tau - 2 (x_{1}-x_{0})   \nonumber \\
  F_1   &=& -\tau -(x_{2}+x_{0}-2x_{1})   \nonumber \\
  F_2   &=& -\tau -(x_{3}+x_{1}-2x_{2})   \nonumber \\
  . &=& . \nonumber \\
  F_{n-1}    &=& -\tau- (x_{n}+x_{n-2}-2x_{n-1})  \nonumber \\
  F_n   &=& -\tau- f(x_n)  - (x_{n+1}+x_{n-1}-2x_{n})  \nonumber \\
  F_{n+1}   &=& -\tau - (x_{n+2}+x_{n}-2x_{n+1}) \nonumber \\
  . &=& . \nonumber \\
  F_{n+m-1}    &=& -\tau- (x_{n+m}+x_{n+m-2}-2x_{n+m-1}) \nonumber \\
  F_{n+m}   &=& -\tau- f(x_{n+m})  - (x_{n+m+1}+x_{n+m-1}-2x_{n+m})  \nonumber \\
  F_{n+m+1}    &=& -\tau- (x_{n+m+2}+x_{n+m}-2x_{n+m+1}), \label{ZRT}
\end{eqnarray}
and in principle the series of equations repeats up to the chain
boundaries with increment of subscripts. We assume that the
mechanical equilibrium is satisfied for all nodes $j$ situated
in between obstacles. Then we have $F_j=0$
but for $j \in  [n,n+m,..n+pm]$. For the segment $j\in[0,n]$, it
is easy to show by recurrence that : $x_j-x_0=-\tau j^2/2$. Thence
the chain shape is parabolic between $n$ and $-n$. For
$j\in[n,n+m]$, we proceed the same and
find $x_{j+n}-x_n=-\tau j(n+j/2) - [F_n+f(x_n)] j$ which
fixes the segment end to \be x_{n+m}=x_n -\tau m(n+m/2)- m
[F_n+f(x_n)].\ee The same can be iterated once again for
$j\in[n+m,n+2m]$ which leads
to $x_{n+2m}-x_{n+m}=-\tau m(n+3m/2) -
(F_n+F_{n+m}+f(x_{n})+f(x_{n+m})) m$. The set of equation on the
positions $x_{n+jm}$ is then:
\begin{eqnarray}
x_{n+m}&=&x_n -\tau m(n+m/2)-  f(x_n) m  \nonumber \\
x_{n+2m}&=&x_{n+m}-\tau m(n+3m/2) - [F_n+F_{n+m}+f(x_n)+f(x_{n+m})] m  \nonumber \\
  . &=& . \nonumber \\
  . &=& . \nonumber \\
x_{n+pm}&=&x_{n+(p-1)m}-\tau m(n+(2p-1)\frac{m}{2}) - m \sum_{j=0}^{p-1} [F_{n+jm}+f(x_{n+jm})]  . \label{ZRT2}
\end{eqnarray}
Subtracting the two latest equations yields: \be \label{try}
F_{n+pm} = -\frac{\Delta_m  x_{n+pm}}{m} -\tau m - f(x_{n+pm}) \ee
where $\Delta_p x_{n+pm}=(x_{n+(p+1)m} + x_{n+(p-1)m} -
2x_{n+pm})$ is the discrete Laplacian applied to the p subscript.
When the entire chain is at mechanical equilibrium $F_{n+pm}=0$
for all $p$. Far enough from the $n_v$-VSC (i.e., the VSC with $n_v$ vacant sites), the solution for
$x_{n+pm}$ tends asymptotically to a constant $x_\infty$ such as
$\tau m = -f(x_\infty)$ and therefore: \be\label{Xinfty} x_\infty
= \frac{2}{\sqrt{3}} \cos(\frac{\arccos(-\tau m
/f_M)}{3}+\frac{4\pi}{3}). \ee We can expend linearly Eq.
\ref{try} for the far enough sites such as the displacement
$x_{n+pm}$ writes $x_{n+pm} = x_\infty +\epsilon_p$ and $
f(x_{n+pm})=f(x_\infty) + f'(x_\infty) \epsilon_p$. Then, at the equilibrium
Eq. \ref{try} yields $[\Delta_p \epsilon_p = -f'(x_\infty) m
\epsilon_p]$ and thence $\epsilon_p$ is an exponential function:
$[\epsilon_p = \epsilon_0 \exp(-\alpha p)]$ which the exponent $\alpha$ verifies \be
\label{eq:alpha}\alpha = \pm 2\ash(\sqrt{-f'(x_\infty)m}/2).\ee
Since the chain displacement is bounded, we are solely concerned
with solutions such as $(\alpha p)>0$.
The sum of the whole set of equations
in Eq. \ref{ZRT2} provides another relation between $\tau$ and the
nodes position $x_{n+pm}$, on the condition that $F_{n+pm}=0$
for all $p$:
\be\label{eq:16} x_{n}-x_{n+pm}=m[\tau
p(n+pm/2)+\sum_{j=0}^{p-1}(p-j)f(x_{n+jm})], \ee
which after
expanding $f(x_{n+jm})$ as a Taylor series around $x_\infty$ and keeping only
the terms linear in p
provides us with an equation which relates $\tau$ to $\epsilon_0$:
\be
\tau= \frac{-\epsilon_0}{(n-m/2)}[ \frac{f'(x_\infty)}{(1-e^{-\alpha})}+\frac{\epsilon_0 f''(x_\infty) }{2(1-e^{-2\alpha})}+\frac{\epsilon_0^2 f'''(x_\infty) }{6(1-e^{-3\alpha})}]\label{toto}.
\ee
The critical chain configuration is reached
when the Hessian associated with Eq. \ref{try}
has a singular eigenvalue. This
allows us to determine the critical value for $\epsilon_0$.
Actually we found that finding the Hessian singular eigenvalue is
equivalent to find the maximum of Eq. \ref{toto} for $\tau$ with respect to $\epsilon_0$.
The solution for the critical bulge is then:
\be\label{ert}
\epsilon_0= -3(1-e^{-3\alpha}) \frac{f''(x_\infty)-\sqrt{f''(x_\infty)^2-4
f'''(x_\infty)f'(x_\infty)\frac{(1-e^{-2\alpha})^2}{3(1-e^{-3\alpha})(1-e^{-\alpha})} } }{2 f'''(x_\infty)(1-e^{-2\alpha})}.
\ee
Combining the solutions for Eqs.
\ref{Xinfty}, \ref{eq:alpha}, \ref{toto} and \ref{ert} allows us to determine the
maximum pinning force associated with $N_m$, the number of obstacle in the denser row.
For this reason, we denote such a maximum as $\tau(N_m)$.
The set of equations giving $\tau(N_m)$ can be solved recursively.
Starting with a small enough trial solution for
$\tau=\tau_0$, we compute the corresponding quantities
$x_\infty$ and $\alpha$ from Eq.
\ref{Xinfty} and from Eq. \ref{eq:alpha}.
Then
$\epsilon_0$ is derived from Eq. \ref{ert} and the corresponding value of $\tau$ from
Eq. \ref{toto}.
If the so obtained quantity is larger than the
initial value $\tau_0$ then the latter is incremented and we
proceed the same up to find identical values for $\tau$ and
$\tau_0$. The end result gives the required $\tau(N_m)$ to a precision fixed by
the trial solution increment.

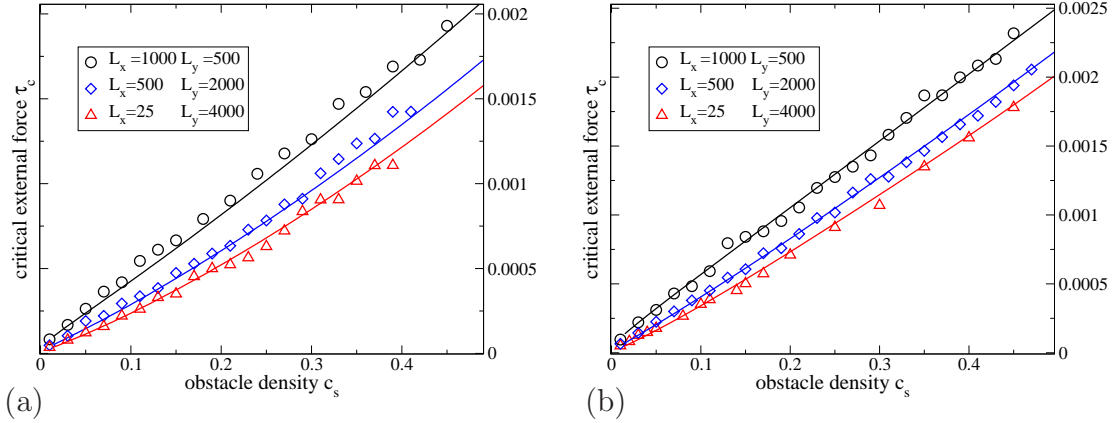
\begin{figure}
\noindent
\includegraphics[width=7.cm]{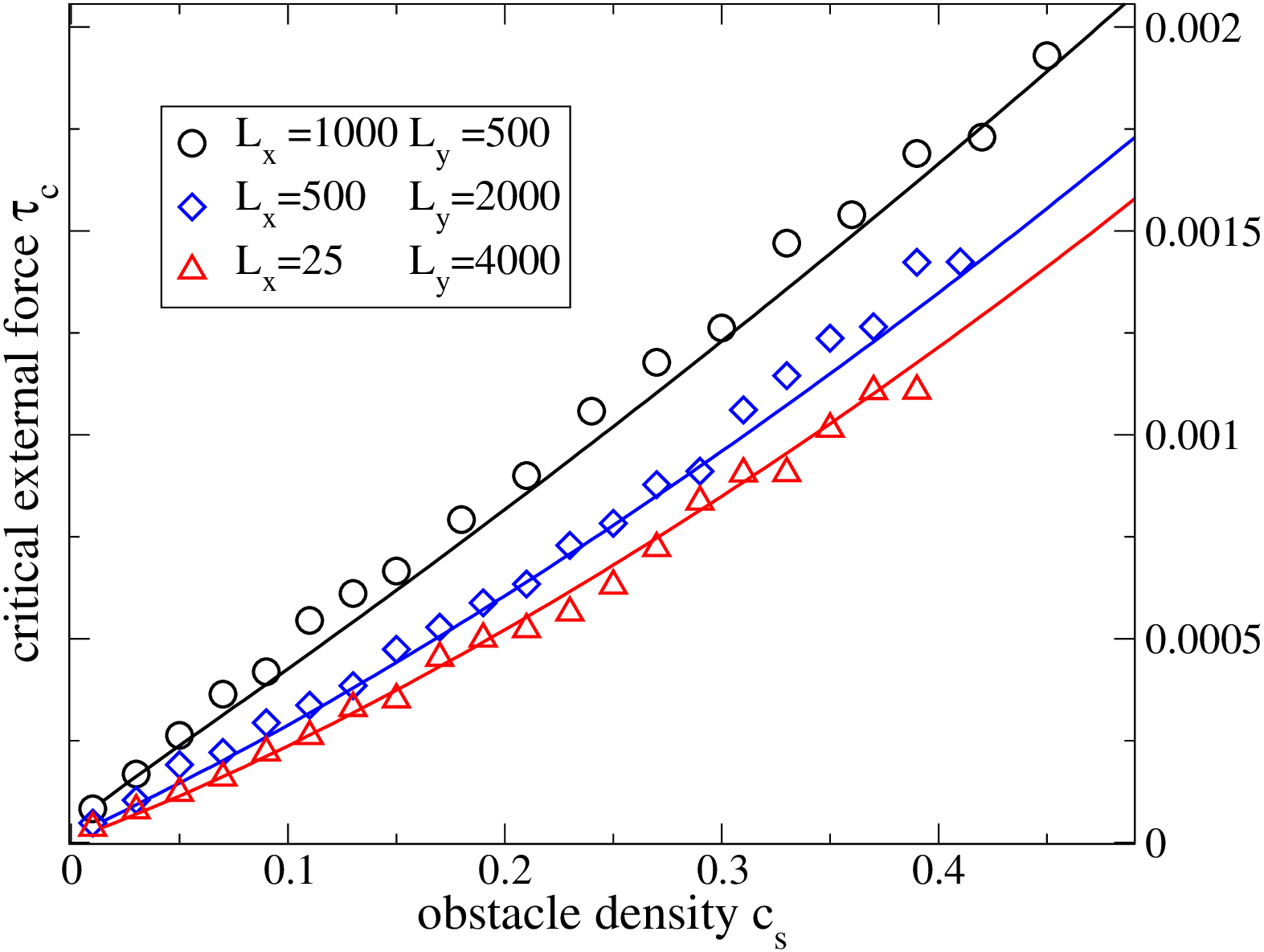}\hspace{0.5cm}
\includegraphics[width=7.cm]{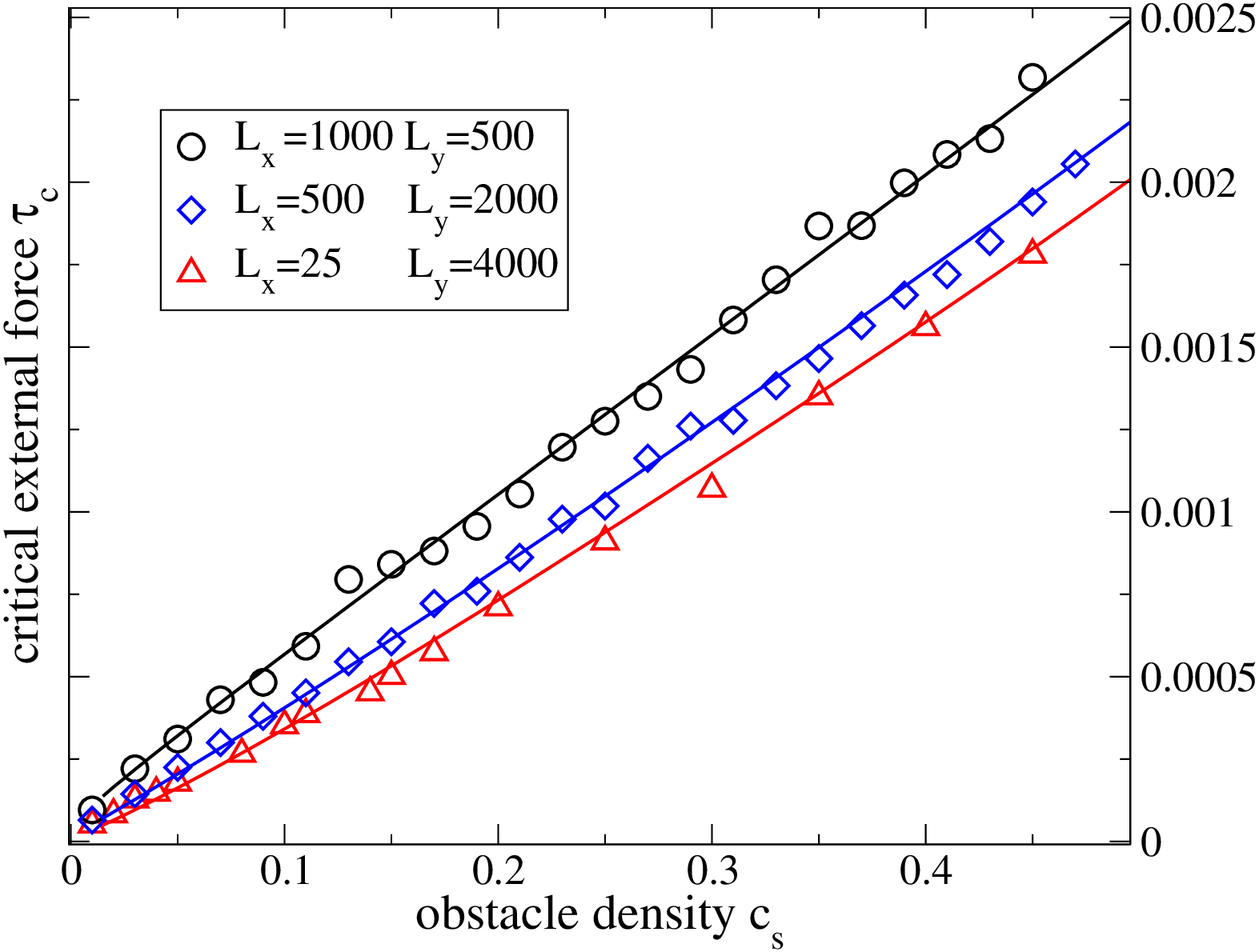}
\caption{\label{HoleSampling} \scriptsize (color online) External
force required to drag a spring chain of length $L_y$ over a
distance $L_x$, for a pinning strength $f_M=0.005$ and a cutoff
$w=0.1$ in (a) and $w=0.5$ in (b). The symbols represent the
data obtained through the simulations described in Sec. \ref{Sec:simul},
for different lattices (see figures legend). The
continuous lines correspond to the predictions made through the
theory detailed in Sec. \ref{HS} for same parameters
as those used in simulations. Colors of symbols and lines correspond
one to one.
}\begin{picture}(100,10)(0,0) \put(5,89){\makebox(0,0){(a)}}
\put(225,89){\makebox(0,0){(b)}}
\end{picture}
\end{figure}
\begin{figure}
\includegraphics[width= 8cm]{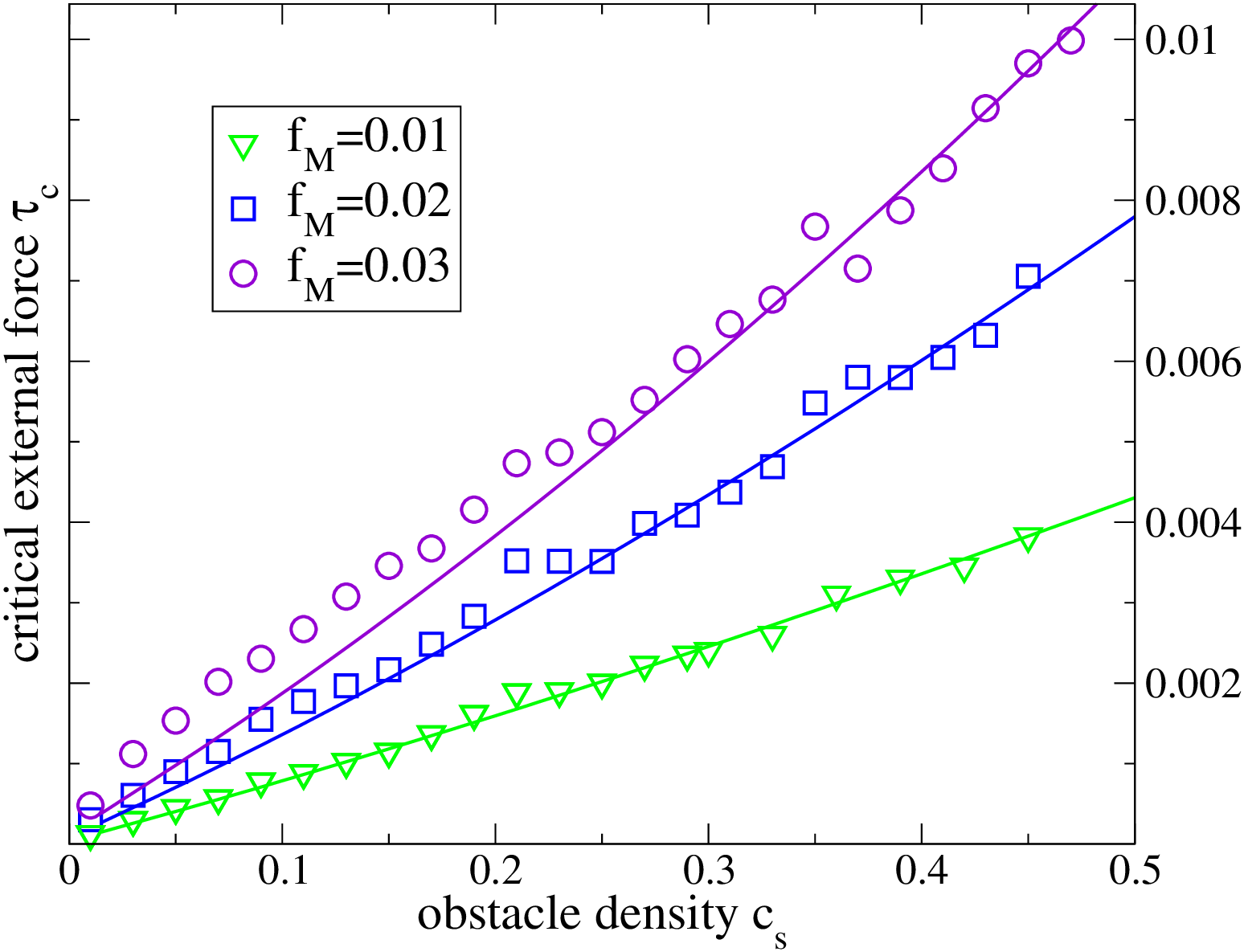}
\caption{\label{HS2} \scriptsize (color online) External force
required to drag a spring chain of length $L_y=1000$ along a distance $L_x=100$,
for different obstacle
pinning strengths (see legend) and the same interaction range $w=0.5$.
The different symbols represent the simulations data and
the continuous lines correspond to the prediction made through the
analytical theory detailed in Sec. \ref{HS} for same parameters.
}\end{figure}

The critical pinning force $\tau_c$ of the random lattice is approximated by
averaging $\tau(N_m)$ over $N_m$:
\be
\tau_c = \sum_{N_m} \beta(N_m) \tau_c(N_m),
\ee
where $\beta(N_m)$ has been given in Eq. \ref{eq:beta}.
The previous theory is compared to
simulations data in Figs. \ref{HoleSampling} (a) and (b) and in
Fig. \ref{HS2} for different lattice dimensions, different pinning
forces, varying $w$ and $f_M$.
A quantitative agreement has
been obtained between theory and simulations, although
no adjustable parameters are involved.
According to same type of
comparisons but for larger $f_M$ values, the previous
analytical work proves relevant for $f_M$ smaller than roughly $0.03$.

\begin{figure}
\includegraphics[width=8.cm]{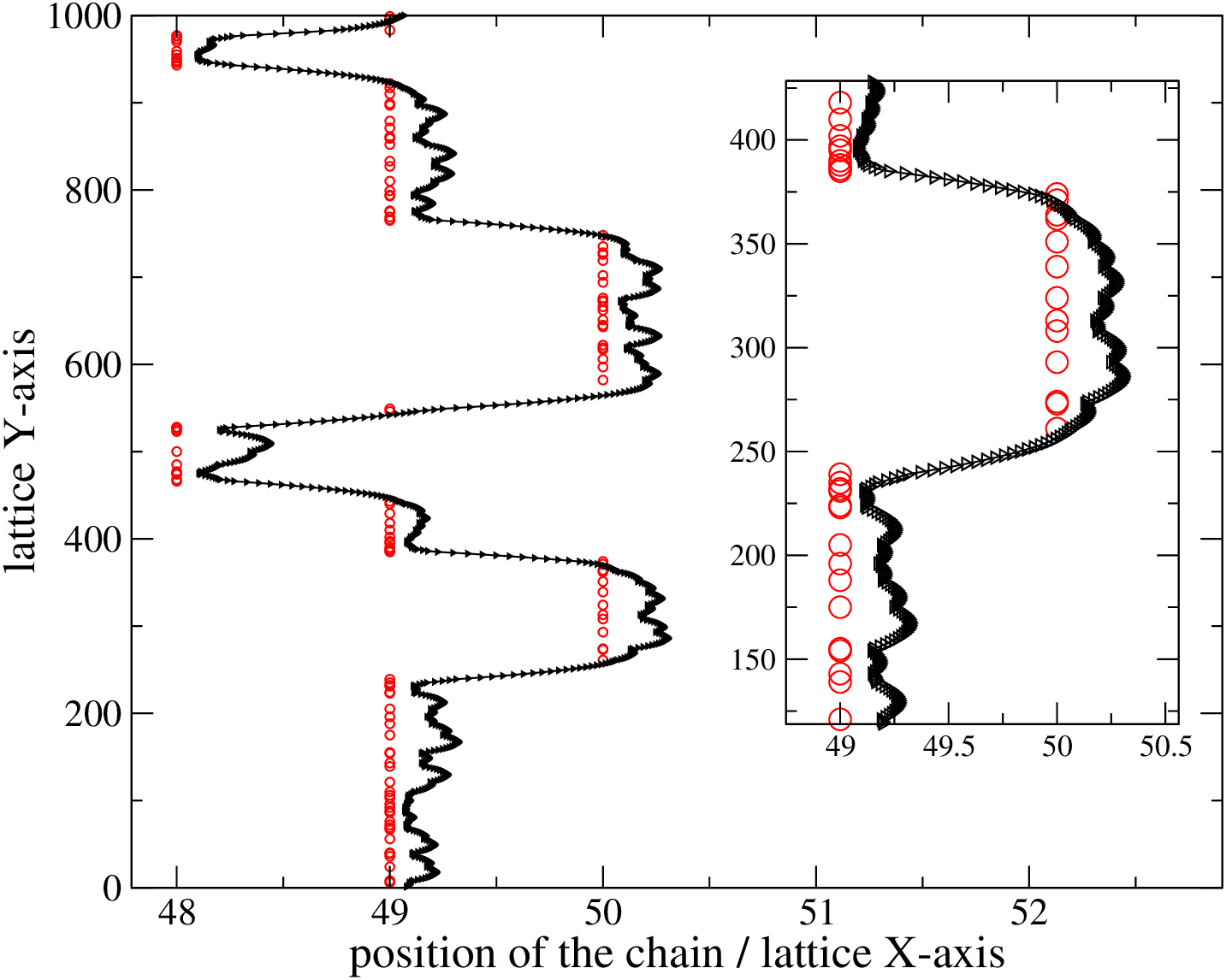}
\caption{\label{Fig:RTY} \scriptsize  (color online)
Strongest pinning configuration for $f_M=0.03$, $L_y=1000$, $L_x=100$, a density
$c_s=9$ $\%$ and an interaction cutoff $w=0.5$. The
inset shows a chain segment the nodes of which are marked with
triangles. Only obstacles close from the string
have been reported as circles.}
\end{figure}

The theory
predictions worsen for cases
where the critical configuration crosses few lattice rows.
As an example, for $f_M = 0.03$ in Fig. \ref{HS2}, the VSC theory is found
to become less accurate for low densities.
In such conditions,
the string
profile at the depinning transition corresponds to a kinked shape,
as shown in Fig. \ref{Fig:RTY}, different from the wavy profile shown in Fig. \ref{fig:FM0_1}
and from the quasi-straight ones shown in Figs. \ref{fLP3} (a) and (b).
From Fig. \ref{HS2}, it can be seen
that the discrepancy increases as the density
decreases while the theoretical predictions remain
accurate for more concentrated obstacle distributions.
As $f_M$ increases above $0.03$, the deviation between theoretical predictions and simulations data
is shifted toward higher densities. In this range of parameters,
the system undergoes a bifurcation, not treated in the present work.

Other authors \cite{Foreman,Rodney2006} noticed that the maximum pinning force
of a random lattice was
dependant of the drag distance.
Concerning the weak pinning points studied here, it is thus of some interest to
explore the variation of the
critical drag force with
lattice dimensions.
The critical drag force was found to vary proportionally
to $[\ln(L_x)]^{\alpha_x}$ where $\alpha_x$ varies with all parameters.
For instance
when $w=0.5$, $f_M=0.01$ and $L_y=1000$, we found $\alpha_x=[0.45-0.036 \ln(c_s)]$.
One can thus conclude that
the $L_x$ dependence is very weak since a fractional power of a
logarithm is a rather wise function.
The maximum drag force depends not only on $L_x$ but also on $L_y$.
The critical drag force varies proportionally to a constant plus
the function $[\ln(L_y)]^{-\alpha_y}$ where $\alpha_y$ decreases
when $c_s$ increases. For instance
when $w=0.5$, $f_M=0.01$ and $L_x=4000$, we found $\alpha_y=[0.26 -0.63 \ln(c_s)]$.
According to direct simulations, the $\tau_c$ dependence in $L_y$ seems to
weaken when $f_M$ becomes large enough for the critical configuration to differ
from the quasi-straight line
as in Fig. \ref{Fig:RTY} or  Fig. \ref{fig:FM0_1}.

For fixed
lattice dimensions, the adjustment of a density power law for the
theoretical critical drag force allows to establish some comparison with
the SSH analytical theory. The density power law fit
is found very close from a linear
variation and may even overpass slightly
the unitary exponent in some situations depending on the lattice geometry.
For instance, from Fig. \ref{HoleSampling} (b), we worked out by
a curve fitting with the form $\tau_c \propto (c_s^\eta)$, an exponent $\eta=1.13$ for
$L_y=4000$ and $L_x=25$ whereas for $L_y=500$ and $L_x=1000$,
$\eta=0.946$ was obtained.
The effective density exponent $\eta$ is therefore size dependent.
A density exponent close from unity
corresponds to the FMS theory for the
hardening of concentrated solid solutions. The CRSS linear
dependence in the solute concentration was also noted in atomistic
calculations on the model solid solution Ni(Al) \cite{Patinet2008}.

\subsection{Extension to broader cutoff}
\label{Sec:HS_B}

To extend the theory developed previously for a short cutoff $w$,
we first consider the case $w \approx 1$, which allows us to
treat the spring chain interaction with only two rows:
the row at the back
and the next nearest one. Such a critical configuration can be seen from direct simulations
as reported in Fig. \ref{fLP3} (b).
To compute $\tau_{c}$, we still employ the same model as depicted in \fig{fLP1}.
However, the interaction with the next nearest row
may weaken the pinning on the back row.
One ascribes to each site $j$ of the next nearest row,
a force $g(x_j)=c_v f(x_j-1)$ where $c_v$ is the density of
obstacles in the next nearest row.
For the situation shown in \fig{fLP1}, the force balance sheet writes as follows:
\begin{eqnarray}
  F_0  &=& -\tau -g(x_0)- 2 (x_{1}-x_{0})   \nonumber \\
  F_1   &=& -\tau -g(x_1)-(x_{2}+x_{0}-2x_{1})   \nonumber \\
  F_2   &=& -\tau -g(x_2)-(x_{3}+x_{1}-2x_{2})   \nonumber \\
  . &=& . \nonumber \\
  F_{n-1}    &=& -\tau-g(x_{n-1})- (x_{n}+x_{n-2}-2x_{n-1})  \nonumber \\
  F_n   &=& -\tau -g(x_{n}) - f(x_n)  - (x_{n+1}+x_{n-1}-2x_{n})  \nonumber \\
  F_{n+1}   &=& -\tau -g(x_{n+1})- (x_{n+2}+x_{n}-2x_{n+1}) \nonumber \\
  . &=& . \nonumber \\
  F_{n+m-1}    &=& -\tau -g(x_{n+m-1})- (x_{n+m}+x_{n+m-2}-2x_{n+m-1}) \nonumber \\
  F_{n+m}   &=& -\tau-g(x_{n+m})- f(x_{n+m})  - (x_{n+m+1}+x_{n+m-1}-2x_{n+m})  \nonumber \\
  F_{n+m+1}    &=& -\tau-g(x_{n+m+1})- (x_{n+m+2}+x_{n+m}-2x_{n+m+1}) \nonumber \\
  . &=& . \label{FFF2}
\end{eqnarray}
To estimate the force exerted by
the next nearest row upon the spring chain,
the function $g(x_j)$ is approximated with a step function:
$g(x_j)=g(x_0)$  if $j\in[0;n-(m-1)/2[$ and
$g(x_j)=g(x_{n+pm})$ if $j\in[n+pm-(m-1)/2;n+pm+(m+1)/2[$.
We then assume that the mechanical equilibrium is achieved for all nodes excepted
those aligned with some obstacles of the back row.
This leads then to
the following set of equations:
\begin{eqnarray}
  \tau  &=& -g(x_0) - 2 (x_{1}-x_{0})   \nonumber \\
  \tau   &=& -g(x_0)-(x_{2}+x_{0}-2x_{1})   \nonumber \\
  . &=& . \nonumber \\
  \tau   &=& -g(x_n)- (x_{n}+x_{n-2}-2x_{n-1})  \nonumber \\
  F_n  &=&-\tau -f(x_n) - g(x_n)  - (x_{n+1}+x_{n-1}-2x_{n})  \nonumber \\
  \tau  &=&  - g(x_n)- (x_{n+2}+x_{n}-2x_{n+1}) \nonumber \\
  . &=& . \nonumber \\
  \tau   &=& - g(x_{n+m})- (x_{n+m}+x_{n+m-2}-2x_{n+m-1}) \nonumber \\
  F_{n+m}&=& -\tau-f(x_{n+m})- g(x_{n+m})- (x_{n+m+1}+x_{n+m-1}-2x_{n+m})  \nonumber \\
  \tau   &=& - g(x_{n+m})- (x_{n+m+2}+x_{n+m}-2x_{n+m+1}), \label{aZRT}
\end{eqnarray}
and the equations repeat up to the
chain boundaries by incrementing subscripts.
By applying recurrence, it is possible to reduce the previous set of equations
to a smaller one, concerning only the regular array of obstacles
in the back row:
\begin{eqnarray}
x_{n+pm}&=&x_{n+(p-1)m}-\tau m(n+(2p-1)m/2)
-m\big[\sum_{j=0}^{p-1} [F_{n+jm}+ f(x_{n+jm})] \nonumber\\
&& + m\sum_{j=0}^{p-2}g(x_{n+jm})  + g(x_0)(n-m/2)\big]
-\frac{m^2-1}{8} g(x_{n+pm}) \nonumber\\
&& - m\frac{3m-1}{4} g(x_{n+(p-1)m}). \label{aZRT2}
\end{eqnarray}
Subtracting the equation for rank ($p-1$) from the one at rank ($p$) gives:
\bq \label{eq:aDelta1}
F_{n+pm}&=& - \Delta_m  x_{n+pm} -\tau m^2 - m f(x_{n+pm})\nonumber\\
&& -\frac{m^2-1}{8}\big[g(x_{n+(p+1)m})+g(x_{n+(p-1)m}) \big] - \frac{3m^2+1}{4} g(x_{n+pm}),
\eq
while for $p=0$:
\bq \label{eq:aDelta1v}
F_{n}&=&   x_{n}-x_{n+m} -\tau m(n+m/2) - m [f(x_{n}) + g(x_0)(n-m/2)]\nonumber\\
&& -\frac{m^2-1}{8}g(x_{n+m})-m\frac{3m-1}{4}g(x_{n}).
\eq
When the equilibrium is achieved $F_{n+pm}=0$ for all p, which allows us to deduce the
asymptotical solution of Eq. \ref{eq:aDelta1} as a constant
$x_{n+pm} \rightarrow x_\infty$ which verifies:
\be
\tau m = -f(x_\infty)-m g(x_\infty).\label{eq:XinftyB}
\ee
The solution to Eq. \ref{eq:XinftyB} is the positive real
root of a third order polynomial equation:
\bq
x_\infty &=& \frac{mc_v}{1+mc_v}+\sqrt{\frac{-2 A_\infty}{3}} \cos{\big(\arccos{\big(\frac{-B_\infty}{2}
\sqrt{\frac{27}{-A_\infty^3}}}\big)/3+\frac{4\pi}{3}\big)}\nonumber\\
&& \text{with} \ A_\infty=\frac{1}{1+mc_v}\big[\frac{3mc_v}{w^2}-1-mc_v-\frac{3(mc_v)^2}{w^2(1+mc_v)} \big]
\nonumber\\
&& \text{and} \ B_\infty=\frac{1}{1+mc_v}\big[\frac{mc_v(\frac{3mc_v}{w^2}-mc_v-1)}{w(1+mc_v)}-2\frac{(mc_v)^3}{w^3(1+mc_v)^2}
\nonumber\\ && \text{\vspace{2.cm} \ \ \ \ \ \ \ \ } +\frac{2m\tau }{3f_M\sqrt{3}}-
\frac{(mc_v)}{w^2}(1/w^2-1) \big]. \label{eq:XinftyBsol}
\eq
If $(1-x_\infty)>w$ then we can set $c_v=0$ in the previous equation which leads to Eq. \ref{Xinfty},
valid for small $w$.
\begin{figure}
\noindent
\includegraphics[width=8.cm]{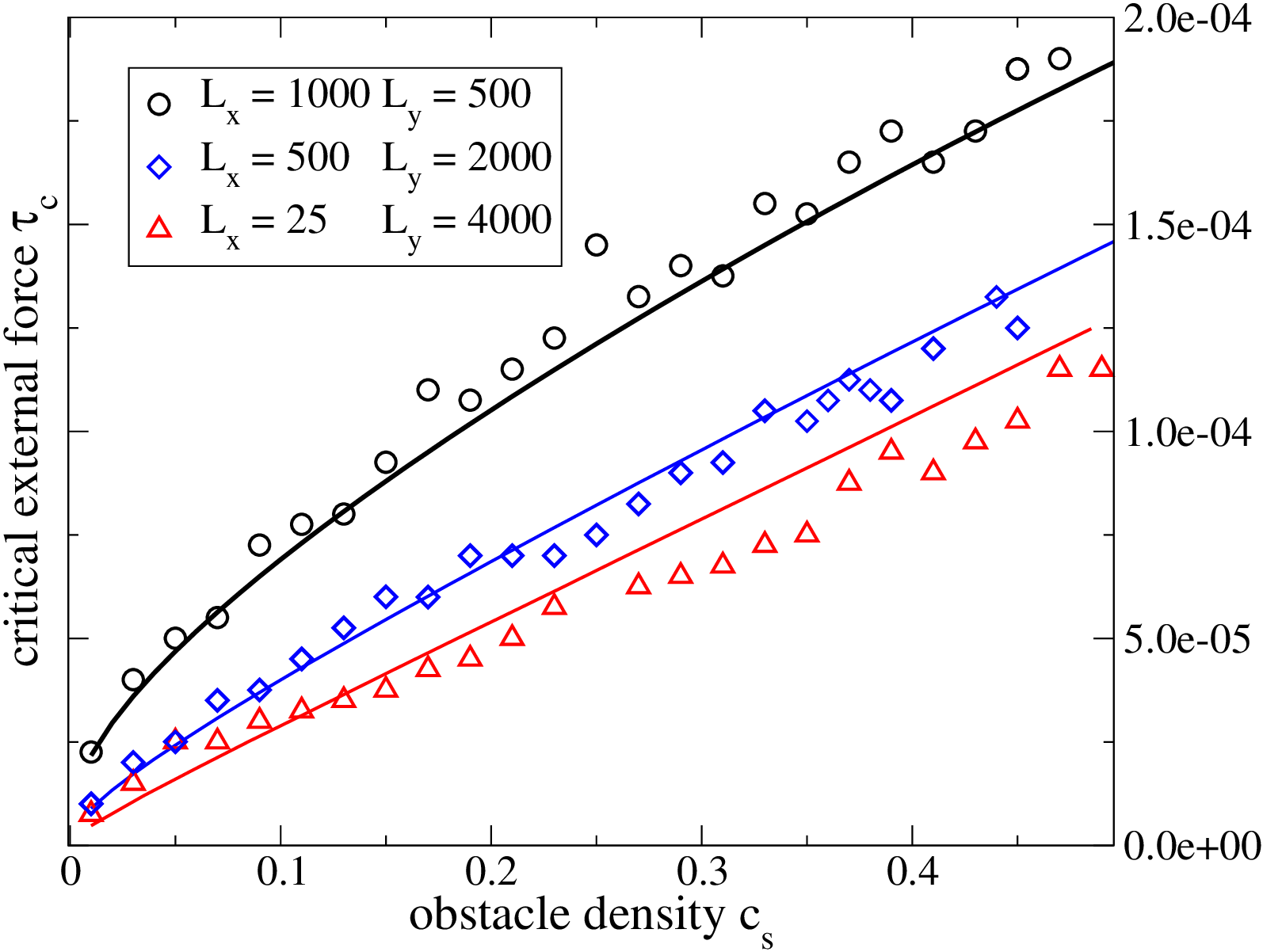}
\caption{\label{HoleSamplingW1} \scriptsize (color online) Same as in Fig.\ref{HoleSampling}
but for a cutoff $w=1$ and a pinning strength $f_M=0.001$.
}\end{figure}
We now expend linearly Eq. \ref{eq:aDelta1}
around $x_\infty$ to determine an approximation of nodes position as
$x_{n+pm}= x_\infty + \epsilon_p $. The equation on $\epsilon_p$ is:
\be \label{eq:aDelta1linear1}
\Delta_1  \epsilon_{p}= - m f'(x_{\infty})\epsilon_p-\frac{3m^2+1}{4} g'(x_{\infty})\epsilon_p-\frac{m^2-1}{8}g'(x_{\infty})
\big[\epsilon_{p+1}+\epsilon_{p-1}\big].
\ee
The solution is an exponential function $\epsilon_p=\epsilon_0 \exp(-\alpha p)$ with
a dispersion relation:
\be\label{eq:alphaROW2ROW}
\alpha=\pm \ach{\big( \frac{2-mf'(x_\infty)-(3m^2+1)g'(x_\infty)/4}{2+\frac{m^2-1}{4}
g'(x_\infty)} \big)}.
\ee
At this stage, it is of some interest to work out the maximum of the spring chain position.
For the segment situated along the larger hole (see Fig. \ref{fLP1}),
$j\in[0,n-(m-1)/2[$, the set of very first equations in Eqs. \ref{aZRT} leads to:
\bq
x_n-x_0&=&-\frac{\tau n^2}{2}
- \frac{m^2-1}{8} g(x_n)-\frac{g(x_0)}{2}[n^2-m^2/4+1/4].\label{eq:X0}\eq
Then $x_0$ can be expressed as a function of $x_n$ since $x_0$ is actually
the positive root of a third order polynomial:
\bq
x_0&=&1+2w\sqrt{\frac{A_0+w}{3A_0}} \cos{\big[ \arccos{\big(-\frac{C_0}{2A_0} \sqrt{\frac{27A_0^3}{(A_0+w)^3}} \big)}/3 +4\pi/3 \big]} \nonumber\\
&& \text{with} \ A_0=\frac{2V_0 c_v}{w}\big[ n^2-m^2/4+1/4\big],\nonumber\\
&& \text{and} \ C_0= x_n-1+\frac{\tau n^2}{2}+\frac{m^2-1}{8}g(x_n).\label{eq:X0b}
\eq
We also need to express the first derivative of $x_0$
against $x_n$ which
according to Eq. \ref{eq:X0} gives:
\bq
\frac{dx_0}{dx_n} = \frac{1+\frac{m^2-1}{8}g'(x_n)}{1-[n^2-(m^2-1)/4]g'(x_0)/2}.\label{eq:dx0}
\eq
The sum of the equations in Eq. \ref{aZRT2} from rank 1 to rank
p, taken sufficiently large, leads to
an equation which the linear term in p is:
\bq\label{toto2}
\tau =&&-g(x_0)+\frac{1}{[n-m/2]}\big[\frac{(m+1)^2}{8m}g(x_\infty)
- \epsilon_0\frac{(f'(x_\infty)+m g'(x_\infty))}{1-e^{-\alpha}}\nonumber\\
&&- \epsilon_0^2\frac{(f''(x_\infty)+m g''(x_\infty))}{2(1-e^{-2\alpha})}
- \epsilon_0^3\frac{(f'''(x_\infty)+m g'''(x_\infty))}{6(1-e^{-3\alpha})}\big].
\eq
The latter equation is similar to Eq. \ref{toto} obtained
for the tightly bound chain, but includes
the interaction with the next nearest row.
The maximum $\tau$ in Eq. \ref{toto2} against $\epsilon_0$ corresponds
to the critical strength which provides a transcendental equation
on $\epsilon_0$:
\bq\label{toto3}
[n-m/2]g'(x_0)\frac{dx_0}{dx_n} =&&
-\frac{(f'(x_\infty)+m g'(x_\infty))}{1-e^{-\alpha}}- \epsilon_0\frac{(f''(x_\infty)+m g''(x_\infty))}{(1-e^{-2\alpha})}
\nonumber\\
&&- \epsilon_0^2\frac{(f'''(x_\infty)+m g'''(x_\infty))}{2(1-e^{-3\alpha})},\nonumber\\
\eq
where $x_n=x_\infty+\epsilon_0$, $x_0$ and $dx_0/dx_n$ are given
in Eq. \ref{eq:X0b} and Eq. \ref{eq:dx0}.
This complete
our computation of the maximum pinning force associated with $n_v$ and $c_v$. The corresponding
value of $\tau$ is therefore related to $N_m$ through Eq. \ref{trans} as well as to the number
of obstacles $N_v=c_vLy$. The maximum force
is now considered as a function of both $N_m$ and $N_v$ and it is denoted by $\tau(N_m,N_v)$.
The probability to find a couple of rows
which actually consists of a back row with $N_m$ obstacles
and a front row with $N_v$ obstacles as being the strongest configuration among $L_x$ rows is
written as:
\begin{equation}\label{eq:teta}
\theta(N_m,N_v) = L_x \ [\rho(N_m)\rho(N_v)]\ [1-\sum_{\tau(N_1,N_2)>\tau(N_m,N_v)} \rho(N_1)\rho(N_2)]^{L_x-1},
\end{equation}
where the function $\rho(N)$ is the binomial
given in Eq. \ref{Eq:Binomial}.
Thence the average critical depinning is given by:
\begin{equation}
\tau_c=\sum_{N_m,N_v} \tau (N_m,N_v) \theta (N_m,N_v).
\end{equation}
The same statistical treatment can be applied to the case $w<0.5$ and yields
the same results
as presented in Sec. \ref{Sec:HS_A} since the critical force given in Eq. \ref{toto} is then independent
of $N_v$.

\begin{figure}
\noindent
\includegraphics[width=8.cm]{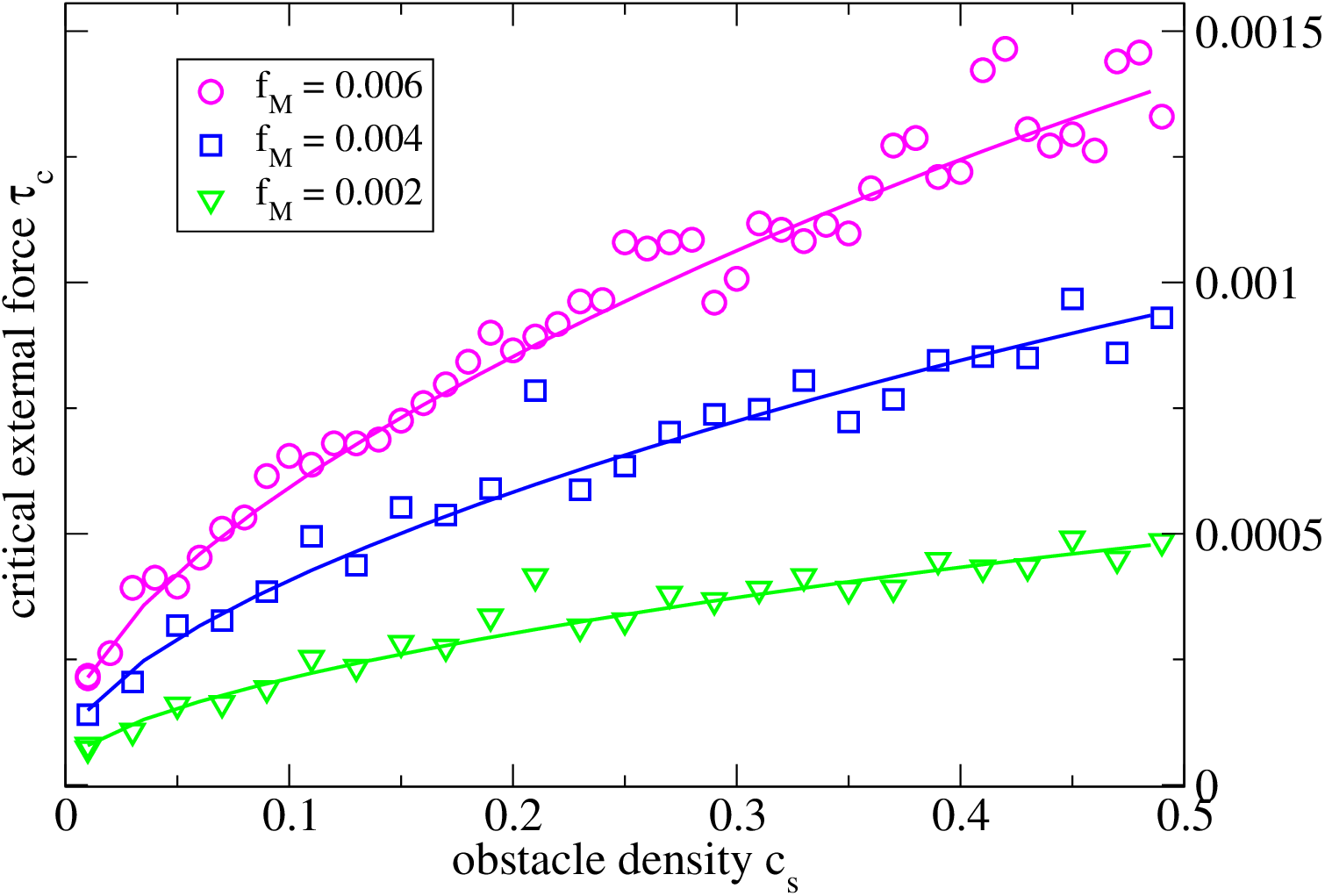}
\caption{\label{HSBC}
\scriptsize (color online) External force
required to drag a spring chain of length $L_y=200$ over a distance $L_x=1500$,
for different obstacle
pinning strengths (see legend) and the same interaction range $w=1$.
The different symbols represent the simulations data and
the continuous lines correspond to the predictions established through the
analytical theory detailed in the text.}
\end{figure}

In Fig. \ref{HoleSamplingW1} and Fig. \ref{HSBC}, the
theoretical predictions are compared to simulations data.
It is worth noticing that the
critical force variation against density
corresponds to a more convex curve than
for shorter cutoff $w\leq 0.5$. An analysis in term of power law fit leads
to an exponent smaller than unity.
For instance in Fig. \ref{HoleSamplingW1}, for $L_y=500$ and
$L_x=1000$, the effective density exponent is around $\alpha=0.65$.
The effective density exponent of $\tau_c$
is found to decrease when $w$ increases (see further Sec. \ref{stillbroader}).
In Fig. \ref{HoleSamplingW1}, the computations
have been performed for different aspect ratios
of the lattice and show the same trend as in Fig. \ref{HoleSampling} for a shorter $w$, the
critical force density exponent increases with
$L_x$ and decreases with $L_y$.
Although a satisfactory agreement is obtained for the different lattices,
we notice that the pinning force $f_M$ is
smaller than in Fig. \ref{HoleSampling}. Actually the field of validity
for the VSC theory is narrower for $w = 1$ than for $w<0.5$.
The limit of application for the theory decreases as the interaction
cutoff increases.
Such a limit also varies with the lattice dimensions: it
decreases when $L_y$ increases and when $L_x$ decreases. For instance, with $L_y=200$ and
$L_x=1500$, the theory proves to be efficient as seen from Fig. \ref{HSBC}
where the comparison has been performed for different pinning forces the maximum
of which is $f_M=0.006$.
The same computations
carried out for $L_y=2000$ and $L_x=500$, keeping constant both $w$ and $f_M$
yield much worse results, in particular for the low densities.
This can be understood comparing the critical profiles for both geometries.
Whereas
the chain profile is quasi-straight for the former
the critical profile is wavy for the latter, i.e., similar to the
one seen in Fig. \ref{fig:FM0_1}.
As for short cutoffs, the system undergoes a bifurcation
passing from a quasi-straight critical profile to a wandering one.
The change in critical profile
occurs for smaller $f_M$ with $w=1$ than for $w<0.5$.


\subsection{Extension to intermediary and still broader cutoffs}
\label{stillbroader}
In previous computations, we developed the pinning force function
as a Taylor series around $x_\infty$, i.e., the asymptotic solution for
the chain nodes position. For intermediary cutoff,
namely $0.5<w<1$,
we must be aware that such a development cannot
be used to approach the non-analytic force function since
the cutoff occurs right in between two rows.
Then $x_\infty$ is assumed to remain far from the
cutoff interaction with the next nearest row. The asymptotic
position $x_\infty$ is thus given by
Eq. \ref{Xinfty} and $\alpha$ by Eq. \ref{eq:alpha}.
The top of the bulge is assumed to be
situated above the cutoff distance from the next nearest row. Then
the equation on $x_0$ is same as in Sec. \ref{Sec:HS_B}
and can be solved analytically as shown in Eq. \ref{eq:X0b}.
The equation relating $\tau$ to $\epsilon_0$ must be rederived.
The sum of the equations in Eq. \ref{aZRT2} from rank 1 to rank
p, taken sufficiently large, leads to
an equation which the linear order in p is now:
\bq\label{toto4}
\tau =-g(x_0)-\frac{1}{[n-m/2]}\big[&&
\epsilon_0\frac{f'(x_\infty)}{1-e^{-\alpha}}+ \epsilon_0^2\frac{f''(x_\infty)}{2(1-e^{-2\alpha})}
+ \epsilon_0^3\frac{f'''(x_\infty)}{6(1-e^{-3\alpha})}\nonumber\\
&+&\sum_{j<l_b} g(x_{n+jm}) \big],
\eq
where $l_b$ is such as $(x_{n+l_b m} > 1 - w)$. The quantity
$l_b$ corresponds to the segment of the bulge which
overpasses the interaction cutoff with next nearest row.
Such equation for $\tau$ holds provided that $l_b$ remains small
compared to $L_y/m$.

\begin{figure}
\noindent
\includegraphics[width=8.cm]{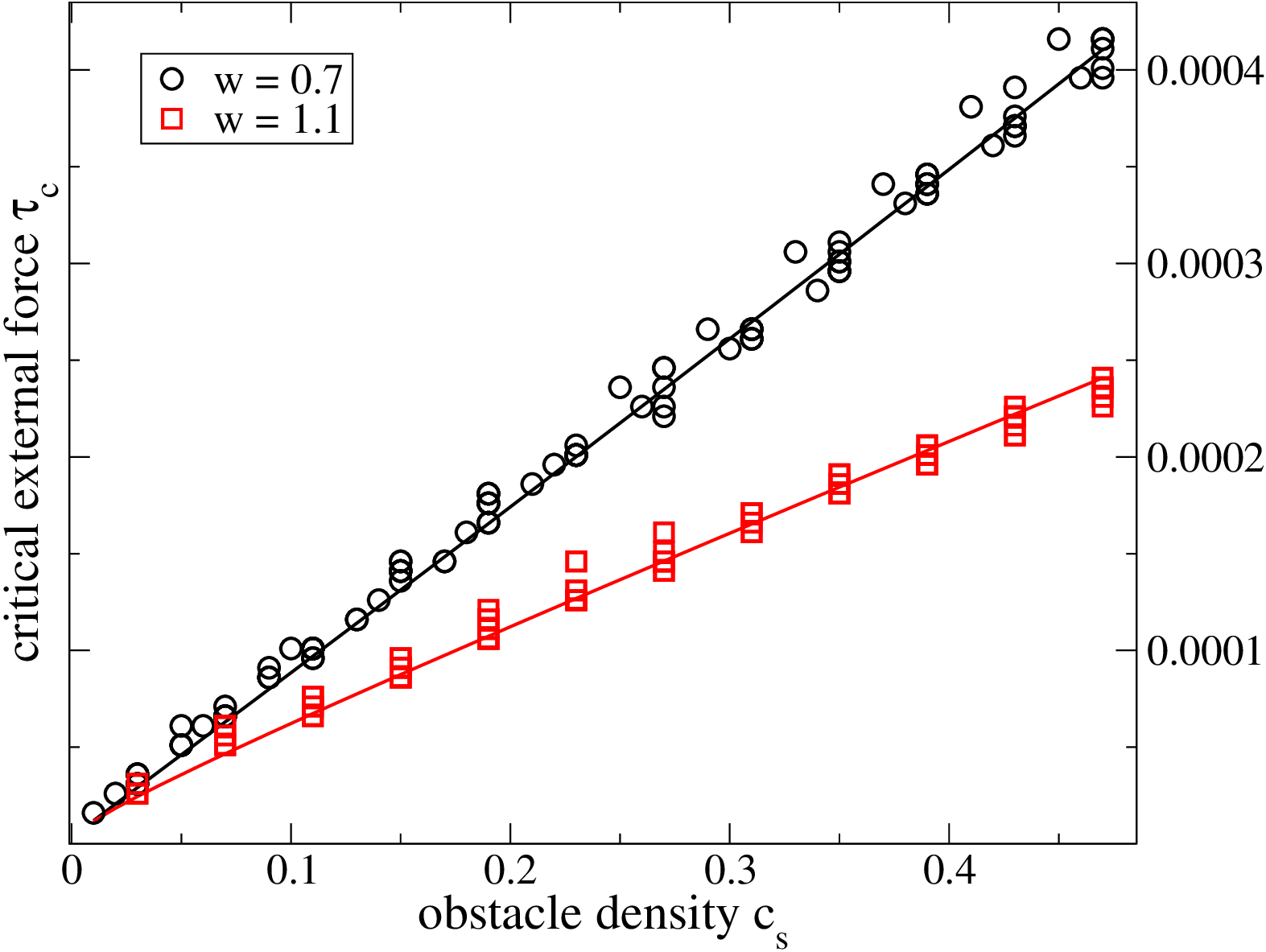}
\caption{\label{HSBC2}
\scriptsize (color online) External force
required to drag a spring chain of length $L_y=1000$ over a distance $L_x=200$,
for a
pinning strength $f_M=0.001$ and different interaction range $w$ (see legend).
The different symbols represent the simulations data and
the continuous lines correspond to the prediction made through the
theory detailed in the text.}
\end{figure}

In Fig. \ref{HSBC2}, the predictions from
the present development are compared to simulations data for $w = 0.7$.
We reported the results obtained from simulations
with different obstacle densities. For some densities,
we carried out the simulations with different
random distributions. The simulations data are distributed around the
theoretical predictions which confirms the robustness of
the theory.

The previous computations can also be extended to
an interaction
cutoff larger than the unit cell parameter (i.e., $w>1$).
In the theory,
the function $g(x_j)=c_v f(x_j-1)$ is substituted with
$g(x_j)=c_v f(x_j-1) + h(x_j)$ where $h(x_j)=c_s f(x_j-2)$ accounts for the
pinning force due to the still further row, i.e., the second next nearest one.
The top of the bulge
may interact with such a row meanwhile
the back of the chain is assumed to remain far from it.
Then the
equations found in Sec. \ref{Sec:HS_B}
for $x_\infty$ and $\alpha$
still hold and $x_0$ is again given by
Eq. \ref{eq:X0} but substituting the function $g(x)$
with the expression $g(x_j)=c_v f(x_j-1) + h(x_j)$.
The equation on $\tau$ follows:
\bq\label{toto5}
\tau =-g(x_0)+\frac{1}{[n-\frac{m}{2}]}\big[&&\frac{(m+1)^2}{8m}g(x_\infty)
- \epsilon_0\frac{(f'(x_\infty)+m g'(x_\infty))}{1-e^{-\alpha}}\nonumber\\
&&- \epsilon_0^2\frac{(f''(x_\infty)+m g''(x_\infty))}{2(1-e^{-2\alpha})}\nonumber\\
&&- \epsilon_0^3\frac{(f'''(x_\infty)+m g'''(x_\infty))}{6(1-e^{-3\alpha})} + \sum_{j<l_b} h(x_{n+jm}) \big],
\eq
where $l_b$ corresponds to the top of the bulge overpassing
the distance cutoff with the second next nearest row,
defined as $(x_{n+l_b m} > 2 - w)$.
In Fig. \ref{HSBC2},
the theoretical computation for $\tau_c$ against the obstacle density
is compared with simulation
data for $(w = 1.1)$ and for different random distributions.
Although the theory could certainly be improved
by accounting for the effect of the nearest row
at the back of the chain, the agreement with simulations data
proves quantitative.

The results obtained in both cases, $w =0.7$ and $w=1.1$ show
that $\tau_c$ decreases with $w$. This was confirmed by other
simulations performed for still larger $w$, up to $w=2$.
Such a decreases is opposite to the one predicted within
the MNL theory for SSH. Moreover according to
the MNL theory, the density exponent of $\tau_c$
is fixed to $\eta= 2/3$, whereas according
to the VSC sampling theory such an exponent depends on
lattice dimensions (see Sec. \ref{Sec:HS_B})
and $w$, as well. For instance, in Fig. \ref{HSBC2},
a power law fit on the
theoretical predictions yields and effective
exponent $\eta = 0.945$ for $w=0.7$
while for $w=1.1$ we obtained $\eta= 0.827$.
The variations $\eta$ with lattice dimensions,
along with its value larger than $2/3$ and its $w$ dependence
demonstrate that the MNL theory does not apply
to the systems studied here. The same conclusion
holds for other
SSH theories predicting a constant density exponent of the critical
depinning threshold.

\section{Summary and perspectives}

In the present study, we addressed a discrete version of a paradigmatic
problem, namely the depinning threshold of an
elastic string on a random substrate. For a planar distribution
of weak pinning
points, a theory was devised to compute the applied force
required to drag the one-dimensional elastic manyfold over a disordered
potential landscape with various aspect ratios.
The theoretical predictions were found accurate provided the critical configuration
remains close from a quasi-straight line bounded between two lattice rows.
The strongest pinning configuration was shown to hinge on the denser
lattice rows in which the largest vacant site clusters (VSC) are bounded in size.
Such maximum VSC correspond to the weakest lattice defects on which
the critical depinning proceeds.
The mean size of the critical VSC is determined through an expression involving only
lattice dimensions and the overall planar obstacle density $c_s$.
The theory allowed us to account for
the finite lattice size effects in a quantitative manner.
The typical variations of the
critical applied force against the chain length and the
drag distance both yield a logarithmic power law.
For a fixed lattice geometry, and an interaction cutoff inferior to half the lattice
cell parameter
the pinning strength was found close to being linear in  $c_s$.
The effective density exponent $\eta$ of the depinning threshold
was found to
depend on the lattice geometry and to
decrease as the interaction cutoff $w$ increases.

In some atomistic studies bearing on dislocation
in a model Ni(Al) solid solution, i.e.,
with a rather marked size effect\cite{Patinet2008},
the maximum force exerted by isolated solute atoms on a dislocation segment
was found
of the order of $0.04$ nano-Newton (nN). The elastic contribution to the
line tension of a screw dislocation, computed
within the isotropic elastic theory \cite{Hirth1982}
is $\Gamma = \Gamma_S \ln(R/ b)$ where $b$ is the Burgers vector,
$R$ is the
inter-dislocation spacing and where the pre-logarithmic factor is
given by $\Gamma_S=\mu b^2 (1+\nu)/4\pi(1-\nu)$,  with $\mu$ as the shear modulus
and $\nu$ as the Poisson coefficient.
The distance $R$ is related to the dislocation density $\rho_d$ by $R=1/\sqrt{\rho_d}$ which
for a standard density in deformed metals $\rho_d = 10^{12}$ m$^{-2}$, gives $R=1$ $\mu$m.
Then, with $\mu_{Ni}=74600$ MPa and $\nu=0.28$,
the screw  dislocation line tension in Ni is $\Gamma=6.1$ nN which gives a ratio between
the obstacle pinning strength and the line tension of
$f_M=0.007$,
of the order of $f_M$ studied in the present work.
A solute atom dilation smaller than the one for Al in a Ni matrix
or a smaller dislocation density
could even yield smaller $f_M$.
In regard of the simplicity of the elastic manyfold model,
the present
study requires to be extended to
a more realistic model involving some of the atomistic details as for instance:
(i) a position dependant obstacle strength,
(ii) a dissociated core, and (iii) mixing of different types of obstacles. Such
extensions would allow to address
longstanding issues in SSH theory.

\bibliographystyle{elsart-num}


\end{document}